\documentclass[fleqn,10pt]{wlscirep}

\usepackage{amsmath}
\usepackage{braket}

\title{Tunable phonon blockade in weakly nonlinear coupled mechanical resonators via Coulomb interaction}

\author[1]{Bijita Sarma}
\author[1, *]{Amarendra K. Sarma}
\affil[1]{Department of Physics, Indian Institute of Technology Guwahati, Guwahati-781039, Assam, India}

\affil[*]{aksarma@iitg.ernet.in}

\begin{abstract}
Realizing quantum mechanical behavior in micro- and nanomechanical resonators has attracted continuous research effort. One of the ways for observing quantum nature of mechanical objects is via the mechanism of phonon blockade. Here, we show that phonon blockade could be achieved in a system of two weakly nonlinear mechanical resonators coupled by a Coulomb interaction. The optimal blockade arises as a result of the destructive quantum interference between paths leading to two-phonon excitation. It is observed that, in comparison to a single drive applied on one mechanical resonator, driving both the resonators can be beneficial in many aspects; such as, in terms of the temperature sensitivity of phonon blockade and also with regard to the tunability, by controlling the amplitude and the phase of the second drive externally. We also show that via a radiation pressure induced coupling in an optomechanical cavity, phonon correlations can be measured indirectly in terms of photon correlations of the cavity mode.
\end{abstract}
\begin{document}

\flushbottom
\maketitle
%
%
\thispagestyle{empty}

\section*{Introduction}
Reaching the quantum regime of micro- and nanomechanical resonators has significance in weak force detection \cite{caves1980measurement, bocko1996measurement} as well as quantum information processing \cite{nielsen2010quantum, stannigel2012optomechanical, rips2013quantum}. Quantum effects can be realized when the mechanical resonator is cooled to its motional ground state, i.e.~when its vibrational energy is higher than or comparable to the thermal noise. Recently, owing to immense progress in the nanofabrication techniques, realizing the quantum regime of a mechanical resonator has become a realistic goal in the field of nano electromechanical and cavity optomechanical systems \cite{schwab2005putting, arcizet2006radiation, gigan2006self, kleckner2006sub, teufel2009nanomechanical, o2010quantum}. Once the quantum regime of a mechanical resonator is reached, it can be further used for quantum information processing related applications. Phonons, which are the quanta of mechanical vibrations, have lower decay rate in comparison to photons. Due to this advantage, phonons have been studied for possible applications in phononic quantum networks \cite{habraken2012continuous, rips2013quantum, gustafsson2014propagating}. In analogy to Coulomb blockade \cite{kastner1993artificial} and photon blockade \cite{imamoglu1997strongly}, it was proposed by Liu {\it{et al.}} that the phonons in a nanomechanical resonator coupled to superconducting charge qubit can exhibit a nonclassical phenomenon called phonon blockade \cite{liu2010qubit}. In such a system, if the nonlinearity is strong enough to give rise to an anharmonic energy level, the excitation of one resonating phonon makes the 
second phonon off-resonant, so that the number of phonons in the
resonator never exceeds one. Phonon blockade based on this mechanism has been studied in a nanomechanical resonator coupled to a qubit \cite{miranowicz2016tunable, wang2016method} or a two-level defect \cite{ramos2013nonlinear}, and also in quadratically coupled optomechanical systems \cite{xie2017phonon, seok2017antibunching},  which requires a strong anharmonicity of the eigenstates corresponding to large coupling strength. Another method has been proposed to obtain phonon blockade in the weak coupling regime via interference of phonon transition pathways \cite{xu2016phonon, guan2017phonon, shi2018tunable}, which is analogous to the unconventional photon blockade effect explored in several systems \cite{liew2010single, bamba2011origin, tang2015quantum, sarma2017quantum, flayac2017unconventional}.

Conventional phonon blockade has been studied in a mechanical resonator with a Kerr-type nonlinearity. \cite{didier2011detecting} The realization of phonon blockade in this system demands strong Kerr-type nonlinearity in order to obtain an anharmonic energy-level. Different from
this, here we show that phonon blockade
in a weakly nonlinear mechanical resonator can be
realized by coupling it to another weakly nonlinear mechanical resonator via Coulomb interaction \cite{hensinger2005ion, ma2014tunable, chen2015dissipation}. Although the nonlinearities in the mechanical resonators are weak, owing to the presence of quantum interference pathways, the system can exhibit phonon blockade. We first discuss the effect of driving only one of the resonators and then driving both the resonators. In the case of single drive, we show that under optimal conditions for the detuning, $\Delta$, and Kerr nonlinearity, $U$, phonon blockade could be achieved. However, by driving both the resonators, one can tune the blockade characteristics by using the optimal values of the drive amplitude and the phase, which can be controlled more conveniently. Also, this modification gives more robustness towards the temperature dependence of the second order correlation function. The detection of phonon blockade by measuring the photon correlations in the presence of an optomechanical interaction is also discussed.


\begin{figure}[!hbt]
	\centering
	\includegraphics [trim={0cm 0cm 0cm 1.2cm},width =0.5\linewidth]{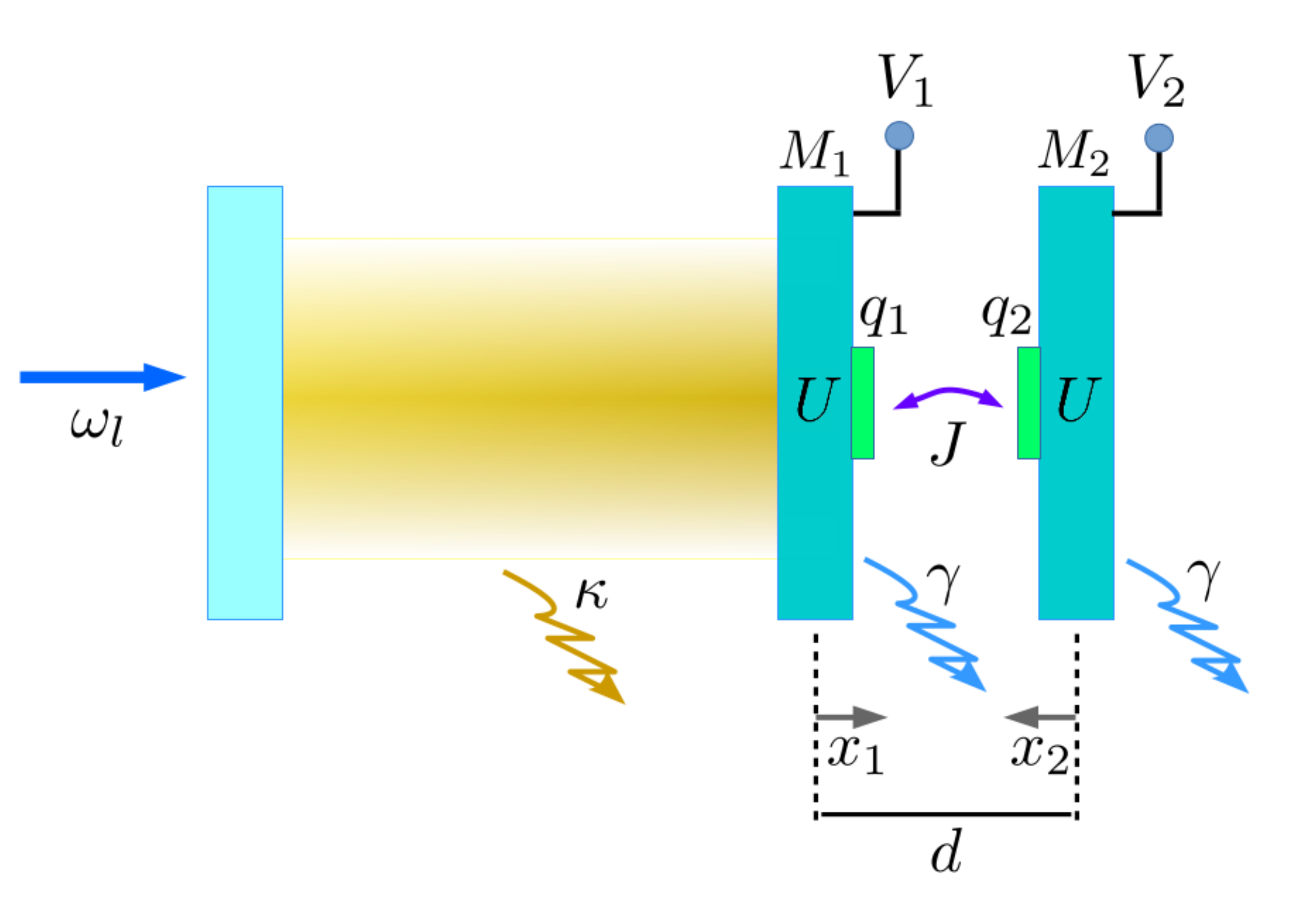}
	\caption {Schematic representation of the optomechanical system, where, $M_1$ is the weakly nonlinear movable end-mirror coupled to another weakly nonlinear mechanical resonator $M_2$ by Coulomb interaction. The electrodes on the resonators carrying charge $q_1$ and $q_2$ are charged by the bias gate voltages $V_1$ and $V_2$. The equilibrium separation of the resonators is $d$. The small deviations of $M_1$ and $M_2$ from their equilibrium positions due to the optomechanical and Coulomb interactions are denoted by $x_1$ and $x_2$ respectively. The cavity linewidth is $\kappa$ and the damping rate of the mechanical resonators is considered to be $\gamma$.}
	\label {fig1}
\end{figure}
\section*{Results}
\subsection*{Model and Hamiltonian}
We consider an optomechanical cavity where the movable end-mirror denoted by $M_1$, is weakly nonlinear, and is coupled to another weakly nonlinear mechanical resonator, $M_2$ via a Coulomb interaction as shown in Fig.~\ref{fig1}. The cavity mode with annihilation operator, $a$, and frequency, $\omega_a$, is driven by a coherent drive with frequency, $\omega_l$. The total Hamiltonian of the system is given by
\begin{align} \label{eq1}
H = H_{\rm{om}}+H_{\rm{m}},
\end{align}
where, $H_{\rm{om}}$ describes the standard linearized optomechanical interaction \cite{ramos2013nonlinear} as given below, with effective optomechanical coupling, $G$, and detuning, $\Delta_a$, in a frame rotating at the drive frequency $\omega_l$: 
\begin{align} \label{om}
H_{\rm{om}}=\Delta_a a^\dagger a + G(a + a^\dagger)(b_1 +  b_1^\dagger) 
\end{align}
The Hamiltonian for the mechanical resonators is given by	
\begin{align} \label{appen1}
H_{\rm{m}}\ =\ H_{\rm{free}} + H_{\rm{nl}} + H_{\rm{co}} + H_{\rm{drive}},
\end{align}
with, $H_{\rm{free}} = \omega_m (b_1^\dagger b_1+ b_2^\dagger b_2)$, $H_{\rm{nl}} = U (b_1^\dagger b_1^\dagger b_1 b_1+ b_2^\dagger b_2^\dagger b_2 b_2)$, $H_{\rm{co}} = \frac{k_e q_1 q_2}{|d+x_1-x_2|}$,\\
and $H_{\rm{drive}} = \Omega_1(b_1^\dagger e^{-i\omega_p t} + b_1 e^{i\omega_p t}) +\Omega_2 (b_2^\dagger e^{-i\phi} e^{-i\omega_q t} + b_2 e^{i\phi} e^{i\omega_q t})$. \\
Here, $b_1$ ($b_1^\dagger$) and $b_2$ ($b_2^\dagger$) are the annihilation (creation) operators for the two mechanical resonator modes with damping rate $\gamma$. Hereafter, we will call the mode `$b_1$' as the primary mode and the mode `$b_2$' as the secondary mode. Here, $H_{\rm{free}}$ is the free Hamiltonian of the two mechanical resonators, $H_{\rm{nl}}$ is the Hamiltonian describing the Kerr nonlinearity, $U$, in both the mechanical resonators and $H_{\rm{co}}$
represents the Coulomb interaction Hamiltonian of the two
charged mechanical oscillators. The primary and the secondary mechanical modes are driven by pumps with frequencies $\omega_p$ and $\omega_q$ respectively with the corresponding pump amplitudes  $\Omega_{1}$ and $\Omega_{2}$ and an initial phase difference $\phi$; which is described by the term $H_{\rm{drive}}$. Hereafter, we will assume that $\omega_q\ =\ \omega_p$.

In the Coulomb interaction Hamiltonian $H_{\rm{co}}$, $k_e$ denotes the electrostatic constant, $d$ is the equilibrium separation of the two charged oscillators in absence of any interaction between them, and $x_1$ and $x_2$ are
the small oscillations of the two mechanical oscillators from their equilibrium positions. Now, assuming that the deviations are small compared to the equilibrium separation, i.e.~$\{x_1,\ x_2\}\ll d$, one can expand	
\begin{align} 
\nonumber
H_{\rm{co}}\ =\ & \frac{k_e q_1 q_2}{d} \left[1-\left(\frac{x_1-x_2}{d}\right)+\left(\frac{x_1-x_2}{d}\right)^2\right].
\end{align}	
Here, the first term is a constant term and the second one is a linear term which can be absorbed into the definition of the equilibrium
positions. The last term consists of two parts: one part refers to the small frequency shift of the original frequencies and can
be neglected by renormalising the mechanical frequencies, and the other part is the coupling term between the oscillators. Therefore, we obtain the Coulomb interaction between the mechanical oscillators as \cite{hensinger2005ion, ma2014tunable, chen2015dissipation}
\begin{align} 
\nonumber
H_{\rm{co}}\ =\ & -\frac{2 k_e q_1 q_2}{d^3} x_1 x_2.
\end{align}	  
The charge contained in the electrodes are given by $q_1 = C_{1} V_{1}$, and $q_2 = -C_{2} V_{2}$, where $C_j$ is the capacitance of the bias gate on the resonator $M_j$. Therefore, $H_{\textrm{co}}$ can be obtained as	
$H_{\rm{co}} = J (b_1 + b_1^\dagger)(b_2+b_2^\dagger)$, where,
$J = \frac{k_e C_1 V_1 C_2 V_2}{d^3}\sqrt{\frac{1}{m_1 m_2 \omega_m^2}}$.
In the weak-coupling regime, considering only the resonant terms, the Coulomb interaction Hamiltonian reduces to 
\begin{align}
\nonumber
H_{\rm{co}}\ =\ J (b_1^\dagger b_2 + b_1 b_2^\dagger).
\end{align}
In the following, we will study the occurrence of phonon blockade in the primary resonator by analyzing the phonon statistics by means of the zero-time delay second-order correlation function given by, $g_b^{(2)}(0)= \langle b_1^\dagger (t) b_1^\dagger (t) b_1(t) b_1(t)\rangle/\langle b_1^\dagger (t) b_1(t)\rangle^2$. 

\subsection*{Phonon blockade with a single drive}
First, we will consider the case when there is no optomechanical interaction. The master equation describing the evolution of the system is given by:
\begin{align} \label{eqmaster1}
\dot{\rho}=i[\rho, H'_m]+\gamma(n_{th,1}+1)L[b_1]\rho+\gamma n_{th,1}L[b_1^\dagger]\rho+\gamma(n_{th,2}+1)L[b_2]\rho+\gamma n_{th,2}L[b_2^\dagger]\rho,
\end{align}
where $L[b_i]\rho= b_i \rho b_i^\dagger -\frac{1}{2} b_i^\dagger b_i \rho -\frac{1}{2}\rho b_i^\dagger b_i$ is the Liouvillian operator for the mode $b_i$ and $n_{th,i}=1/[\exp(\hbar \omega_m/k_B T)-1]$ denotes the thermal phonon number in that mode at environmental temperature $T$. We will consider $n_{th,1}=n_{th,2}=n_{\rm{th}}$ for the rest of the paper.  
The Hamiltonian describing the mechanical resonators in a rotating frame with the mechanical drive frequency is given by 
\begin{align} \label{eqHnew}
H'_m=& \Delta b_1^\dagger b_1+ \Delta b_2^\dagger b_2 + U b_1^\dagger b_1^\dagger b_1 b_1 + U b_2^\dagger b_2^\dagger b_2 b_2+ J(b_1^\dagger b_2 + b_1 b_2^\dagger)+
\Omega_1(b_1^\dagger + b_1) +\Omega_2(b_2^\dagger e^{-i\phi} + b_2 e^{i\phi}),
\end{align}
\begin{figure}[!hbt]
	\centering
\includegraphics [width =0.8\linewidth]{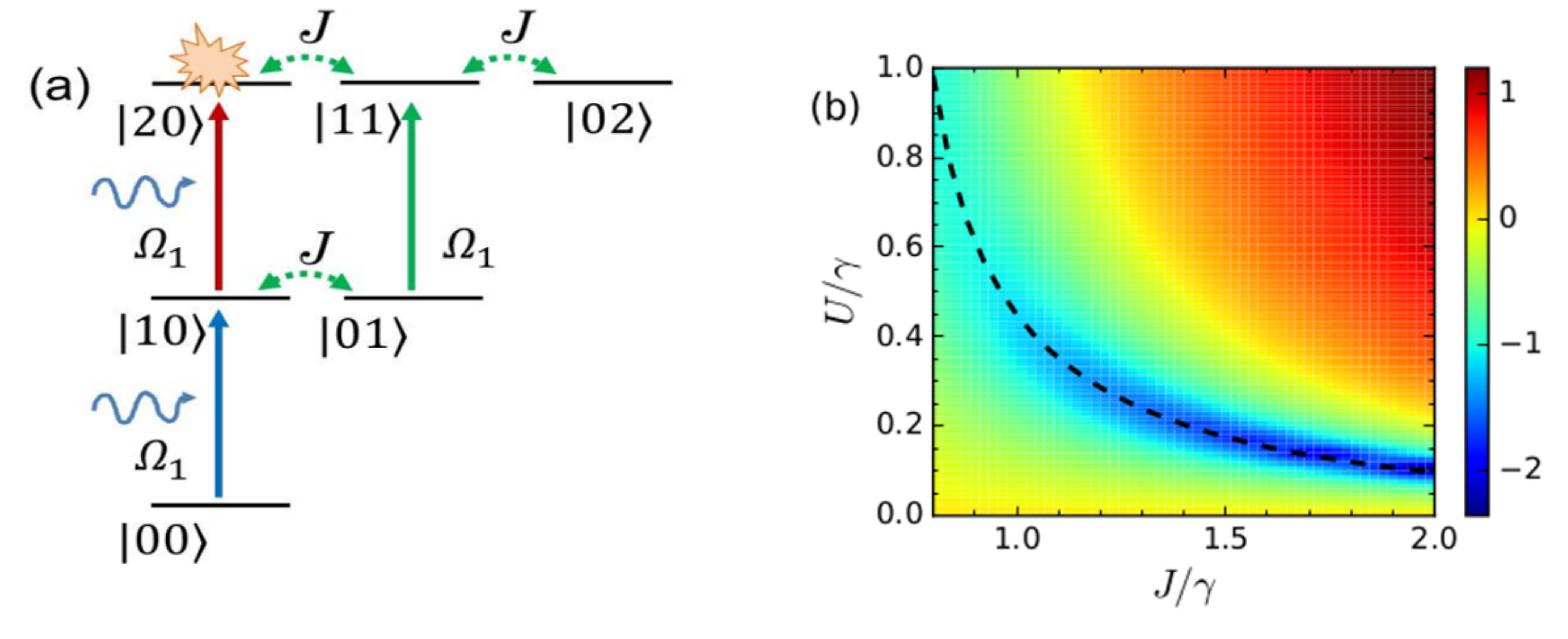}
	\caption {(a) Different two-phonon excitation paths that can lead to interference-based phonon blockade for a single pump applied on the primary resonator, and (b) logarithmic plot of $g_b^{(2)}(0)$ as functions of normalized $U$ and $J$ with optimum values of $\Delta$. The black dashed line represents the optimum values of $U/\gamma$ corresponding to the values of $J/\gamma$.}
	\label {fig2}
\end{figure}
\begin{figure}[!hbt]
	\centering
\includegraphics [width =1.0\linewidth]{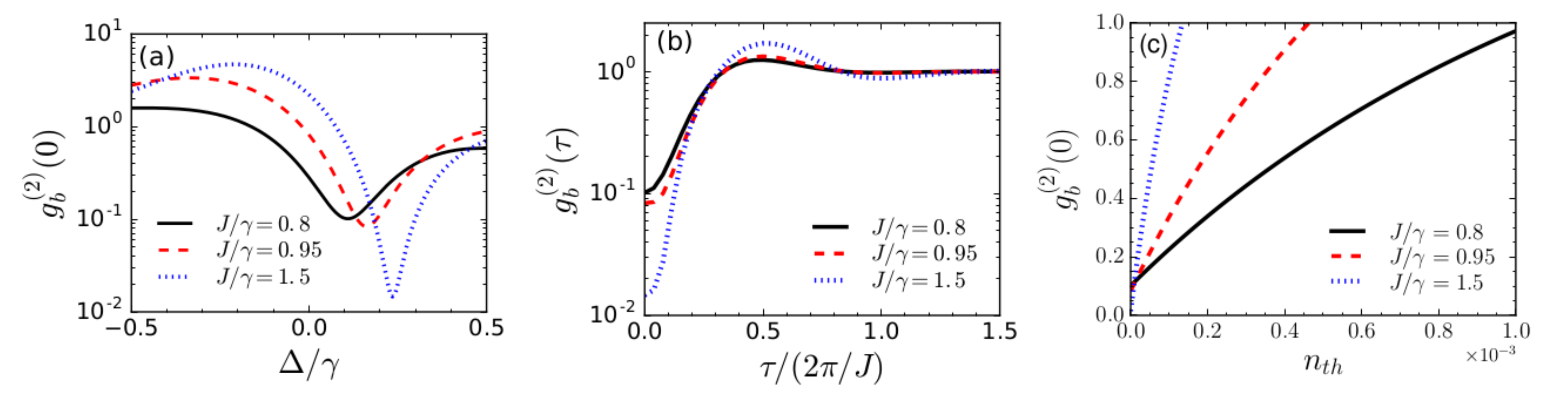}
	\caption {(a) Plot showing the variation of $g_b^{(2)}(0)$ as a function of $\Delta/\gamma$ with $U=U_{\rm{opt}}$ for different values of $J$. (b) Second-order correlation function with finite time-delay, $g_b^{(2)}(\tau)$ and (c) temperature dependence of $g_b^{(2)}(0)$. Other parameters are: $\Omega_1 = 0.1 \gamma$, $U=U_{\rm{opt}}$ and $\Delta = \Delta_{\rm{opt}}$.}
	\label {fig3}
\end{figure}
where, $\Delta = \omega_m - \omega_p$ is the detuning from the mechanical pump frequency. We will calculate $g_b^{(2)}(0)$ numerically by solving Eq.~\eqref{eqmaster1} in the weak-driving limit i.e.~for $\{\Omega_1, \Omega_2\} \ll \gamma$, from $g_b^{(2)}(0) =  \textrm{Tr} (b_1^\dagger b_1^\dagger b_1 b_1 \rho_{\textrm{ss}})/[\textrm{Tr}( b_1^\dagger b_1\rho_{\textrm{ss}})]^2$, where $\rho_{\textrm{ss}}$ is the steady-state density matrix. Before solving the master equation numerically, in order to obtain the optimal parameters for unconventional phonon blockade, we develop an analytical model in the following. Firstly, we consider the case when the secondary mechanical resonator is not driven, i.e.~$\Omega_2 =0$. At low temperature, and assuming a weak pumping condition, the low-energy levels dictated by the Hamiltonian is shown in Fig.~\ref{fig2}(a). Assuming that the system is initially prepared in the $|00\rangle$ state, we consider the following ansatz:
\begin{align} \label{eq4}
|\psi\rangle=&  C_{00}|00\rangle+C_{10}|10\rangle+C_{01}|01\rangle+C_{20}|20\rangle+C_{11}|11\rangle + C_{02}|02\rangle.
\end{align}
The coefficients $C_{ij}$'s can be obtained by solving the Schr\"{o}dinger equation $i\frac{d|\psi\rangle}{dt}=H_{\rm{eff}}|\psi\rangle$, where, $H_{\rm{eff}}= H'_m-i\frac{\gamma}{2} b_1^\dagger b_1-i\frac{\gamma}{2} b_2^\dagger b_2$ is the non-Hermitian Hamiltonian that includes the damping of the mechanical oscillators. Following an iterative method prescribed by Bamba {\it{et al.}} in connection with photon blockade in coupled photonic molecules \cite{bamba2011origin}, in the limit of weak $\Omega_1$, at steady-state, the optimal parameters are obtained as follows:
\begin{align} \label{eqoptimal1}
\Delta_{\rm{opt}}\ =\ \pm \frac{1}{2} \sqrt{\sqrt{9J^4+8\gamma^2 J^2}-\gamma^2-3J^2}\ , \qquad
U_{\rm{opt}}\ =\ \frac{\Delta_{\rm{opt}}(5\gamma^2+4\Delta_{\rm{opt}}^2)}{2(2J^2-\gamma^2)}.
\end{align}	
The limit for the coupling, $J$, in this case is that the value of $J$ must be larger than $\gamma/\sqrt{2}$. In Fig.~\ref{fig2}(b), we show the variation of the zero time-delay second-order correlation function $g_b^{(2)}(0)$ by solving the master equation, i.e. Eq.~\eqref{eqmaster1} , in a truncated Fock space. Here, $g_b^{(2)}(0)$ is plotted as functions of the normalized coupling strength $J/\gamma$ and nonlinearity $U/\gamma$ for $U\leq \gamma$, with optimal values of $\Delta$ as derived in Eq.~\eqref{eqoptimal1}. The black dashed curve shows the optimal values of $U$ calculated in Eq.~\eqref{eqoptimal1}. It is observed that for the optimal conditions, phonon blockade can be obtained in the weakly nonlinear regime.

To demonstrate these results more clearly, in Fig.~\ref{fig3}(a), $g_b^{(2)}(0)$ is depicted as a function of $\Delta/\gamma$ for different values of $J/\gamma$. The value of $U$ is considered to be $U_{\rm{opt}}$. For $J/\gamma = 0.8$, $0.95$, and $1.5$, the optimal values of $\Delta/\gamma$ found from the analytical calculations are $\approx 0.11$, $0.16$, and $0.24$ respectively. The corresponding optimal values of $U/\gamma$ are $0.98$, $0.52$, and $0.18$ respectively. From the plots, it is evident that the numerically calculated results show complete agreement with the optimal values of the parameters calculated from the approximate analytical model. With weak coupling strengths of $J/\gamma =0.8$ and $0.95$, $g_b^{(2)}(0)\approx 0.1$,while for a moderate value of $J/\gamma = 1.5$, $g_b^{(2)}(0)$ is on the order of $0.01$. We also demonstrate the second-order correlation function, $g_b^{(2)}(\tau)= \langle b_1^\dagger (t) b_1^\dagger (t+\tau) b_1(t+\tau) b_1(t)\rangle/ \langle b_1^\dagger (t) b_1(t)\rangle^2$, as a function of the normalized time delay $\tau/(2\pi /J)$ in Fig.~\ref{fig3}(b). Considering optimal parameters, when $J/\gamma = 0.8$ and $0.95$, $g_b^{(2)}(0)\approx 0.1$ at $\tau=0$, and for increasing delay times $g_b^{(2)}(\tau)>g_b^{(2)}(0)$. Similarly, for $J/\gamma = 1.5$, $g_b^{(2)}(0)\approx 0.01$ at $\tau=0$, and for higher delay times $g_b^{(2)}(\tau)>g^{(2)}(0)$ and finally reaches the value $1$. Therefore, the plots demonstrate that the phonons are antibunched and have sub-Poissonian distribution. Now, in order to see the influence of environmental phonon population on the phonon blockade characteristics, in Fig.~\ref{fig3}(c), we show the variation of $g_b^{(2)}(0)$ as a function of the bath phonon number, $n_{\rm{th}}$. For $J/\gamma=0.8$, $g_b^{(2)}(0)$ reaches $1$ at $n_{\rm{th}} \approx 0.001$, whereas, for $J/\gamma = 0.95$ and $1.5$, $g_b^{(2)}(0) \leq 1$ upto $n_{\rm{th}}=4.5\times 10^{-4}$ and $n_{\rm{th}} = 1.5\times 10^{-4}$ respectively. Therefore, it is evident that the environmental thermal population has undesirable effect on the observation of phonon blockade.

\subsection*{Phonon blockade with two drives}
\begin{figure*}[!hbt]
	\centering
	\includegraphics [width =0.35\linewidth]{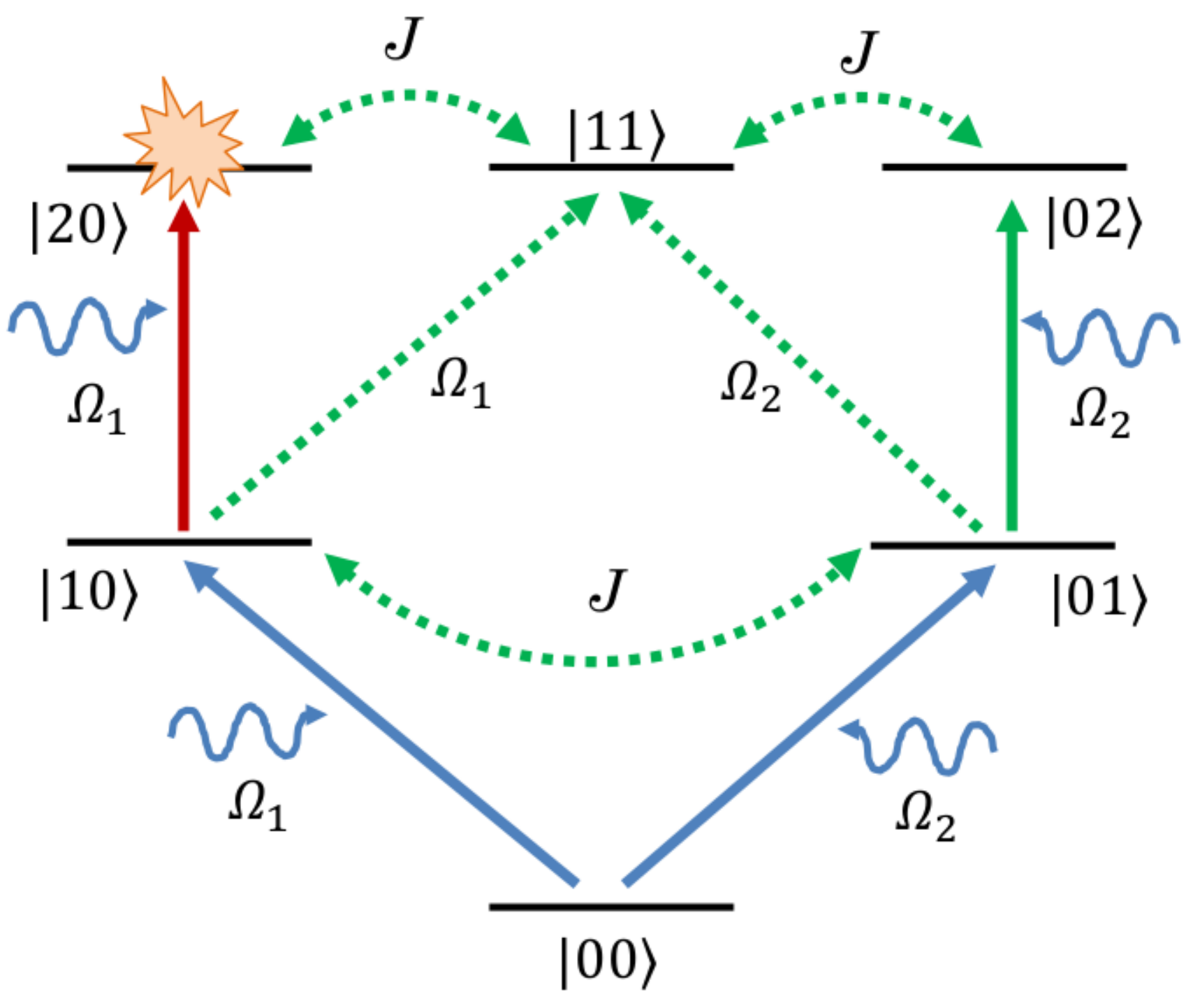}\\
	\caption {Different paths for two-phonon excitation when an additional pump is applied on the secondary mechanical resonator.}
	\label {fig4}
\end{figure*}
\begin{figure*}[!hbt]
	\centering
\includegraphics [width =0.8\linewidth]{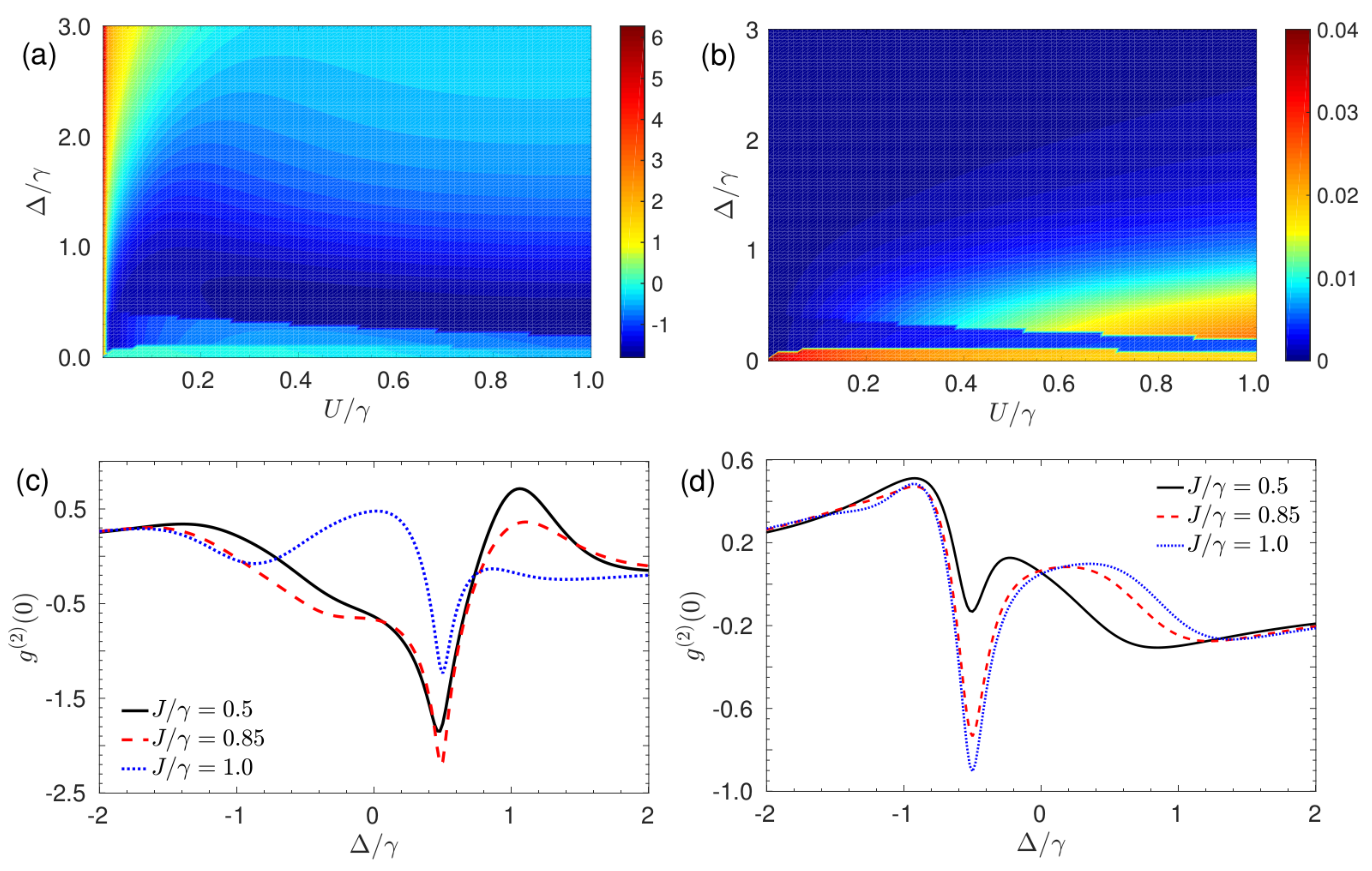}
	\caption {(a) Contour plot showing the variation of zero-time-delay second-order correlation function $g_b^{(2)}(0)$ as function of normalized detuning $\Delta_a/\kappa$ and $U/\kappa$ and (b) phonon number in the primary resonator for $\zeta_+,\ \phi_+$. (c) Variation of $g_b^{(2)}(0)$ for different values of $J$ with $U = 0.5 \gamma$ and $\Delta = 0.5 \gamma$. The black solid, red dashed and blue dotted lines correspond to $J = 0.5 \gamma$, $0.85 \gamma$ and $\gamma$, and the corresponding values of $\{\zeta_+, \phi_+/\pi\}$ are $\{2.59, 0.20\}$, $\{1.60, 0.23\}$ and $\{0.39, 0.06\}$ respectively. (d) shows the variation of $g_b^{(2)}(0)$ for $U_{\rm{opt}} = 0.5 \gamma$ and $\Delta_{\rm{opt}} = -0.5 \gamma$ for $J = 0.5 \gamma$ (black solid line), $0.85 \gamma$ (red dashed line) and $\gamma$ (blue dotted line). The respective values of $\{\zeta_+, \phi_+/\pi\}$ are $\{1.09, 0.88\}$, $\{0.82, 0.93\}$ and $\{0.77, 0.95\}$.}
	\label {fig5}
\end{figure*}
We now turn to study the phonon correlations by applying an additional drive $\Omega_2$ on the secondary mechanical resonator. The transition paths leading to two-phonon excitation, are shown in Fig.~\ref{fig4}. Analytical calculations of optimal conditions in this case gives rise to a quadratic equation in $\zeta e^{-i \phi}$:
\begin{align} 
a_2 \zeta^2 e^{-2i\phi} + a_1 \zeta e^{-i\phi} + a_0\ =\ 0, 
\end{align}
with $\zeta = \Omega_2/\Omega_1$, $a_2 = 2J^2 (\Delta'+U/2)$, $a_1 = -4J \Delta' (\Delta'+U)$, $a_0  = 2 \Delta'^3 + U(J^2 + 2 \Delta'^2)$ and $\Delta' = \Delta-i\frac{\gamma}{2}$. The solutions of the quadratic equation are given by:
\begin{align} \label{eq10}
\zeta_{\pm} e^{-i\phi_{\pm}}\ =\ \frac{1}{J^2 (U+2\Delta')}\left[2 J \Delta' (U+\Delta') \pm \sqrt{J^2 U(2 U \Delta'^2 + 2 \Delta'^3 - J^2 U - 2 J^2 \Delta')} \right].
\end{align}
From Eq.~\eqref{eq10}, it can be seen that for specific values of the parameters, $U$, $J$ and $\Delta$, the optimal values of $\zeta$ and $\phi$ could be obtained, and there are two optimal values of $\zeta$ and $\phi$ for a specific set of system parameters. Therefore, by applying the additional pump we can choose the optimal values of the amplitude and the phase of the second drive for different coupling strengths and detuning in the system.

Fig.~\ref{fig5}(a) depicts $g_b^{(2)}(0)$ as functions of the rescaled detuning $\Delta/\gamma$ and $U/\gamma$ corresponding to $\zeta_{+}, \phi_+$ for a weak coupling value of $J= 0.5\gamma$. In Fig.~\ref{fig5}(b), we show the corresponding average phonon number in the primary resonator. From these plots, it is observed that for the parameter regime where $g_b^{(2)}(0)$ is found to be on the order of $0.01$, average phonon number on the order of $0.01$ could be obtained. We show the variation of $g_b^{(2)}(0)$ as a function of $\Delta/\gamma$, for different values of $J$ in Fig.~\ref{fig5}(c), with $U_{\rm{opt}}/\gamma = 0.5$ and $\Delta_{\rm{opt}}/\gamma = 0.5$, and $J/\gamma = 0.5$, $0.85$ and $1$. It is observed that phonon blockade could be obtained at $\Delta = 0.5 \gamma$, which is in agreement with $\Delta_{\rm{opt}}$, as predicted by the analytical calculations. Fig.~\ref{fig5}(d) shows the variation of $g_b^{(2)}(0)$, with $U_{\rm{opt}} = 0.5 \gamma$, and $\Delta_{\rm{opt}} = -0.5 \gamma$, and it is observed that phonon blockade could be obtained at $\Delta = -0.5 \gamma$.

\begin{figure}[!hbt]
	\centering
   \includegraphics [width =0.8\linewidth]{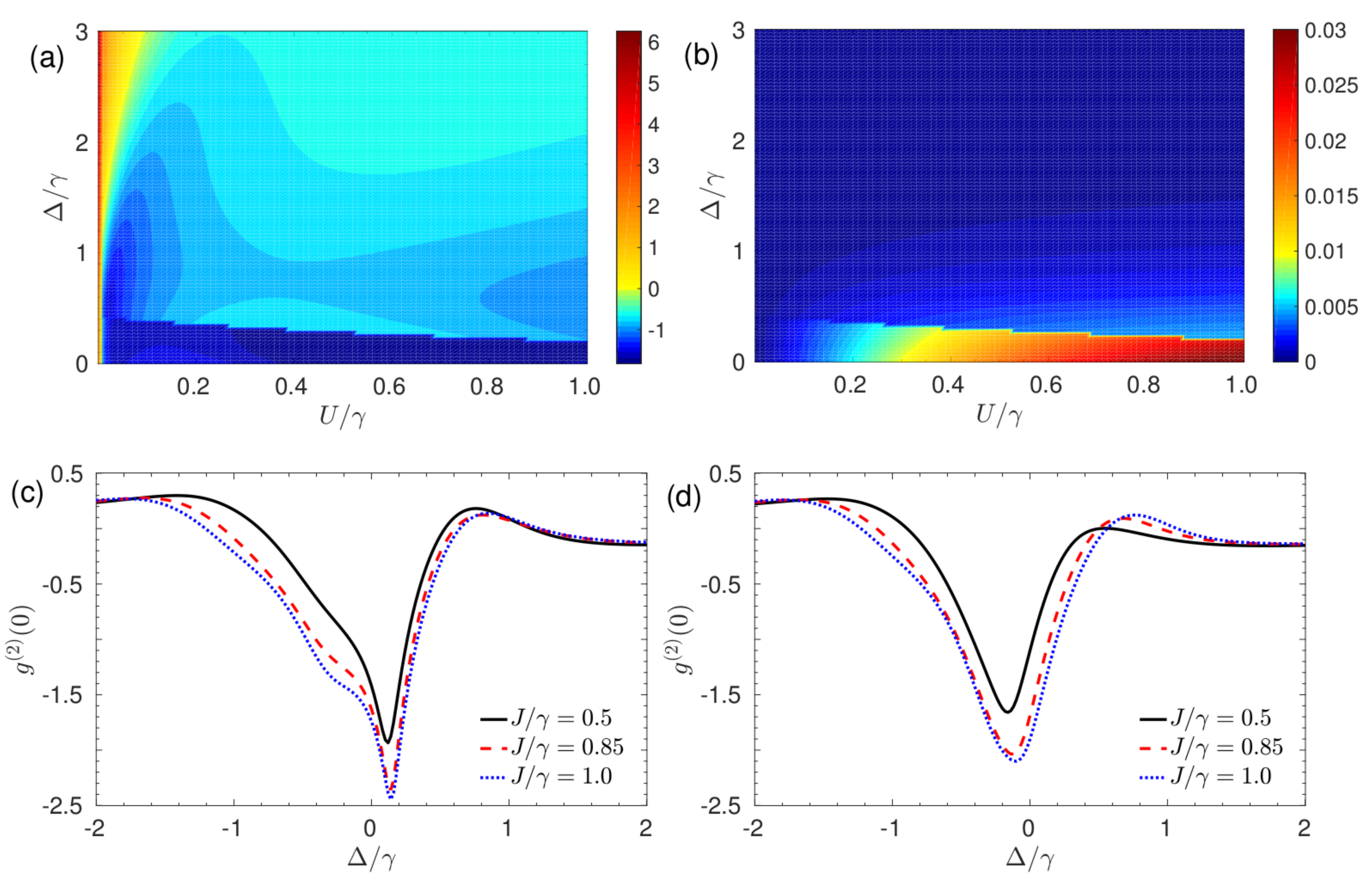}
	\caption {(a) Contour plot showing the variation of zero-time-delay second-order correlation function $g_b^{(2)}(0)$ as function of normalized detuning $\Delta_a/\kappa$ and $U/\kappa$ and (b) phonon number in the primary resonator for $\zeta_-,\ \phi_-$. (c) Variation of $g_b^{(2)}(0)$ for different values of $J$ with $U = 0.5 \gamma$ and $\Delta = 0.15 \gamma$. The black solid, red dashed and blue dotted lines correspond to $J = 0.5 \gamma$, $0.85 \gamma$ and $\gamma$, and the corresponding values of $\{\zeta_-, \phi_-/\pi\}$ are $\{2.25, 0.30\}$, $\{1.51, 0.32\}$ and $\{1.37, 0.33\}$ respectively. (d) shows the variation of $g_b^{(2)}(0)$ for $U_{\rm{opt}} = 0.5 \gamma$ and $\Delta_{\rm{opt}} = -0.15 \gamma$ for $J = 0.5 \gamma$ (black solid line), $0.85 \gamma$ (red dashed line) and $\gamma$ (blue dotted line). The corresponding values of $\{\zeta_-, \phi_-/\pi\}$ are $\{2.23, 0.38\}$, $\{1.52, 0.37\}$ and $\{1.37, 0.36\}$ respectively.}
	\label {fig6}
\end{figure}
\begin{figure}[!hbt]
	\centering
\includegraphics [width =0.8\linewidth]{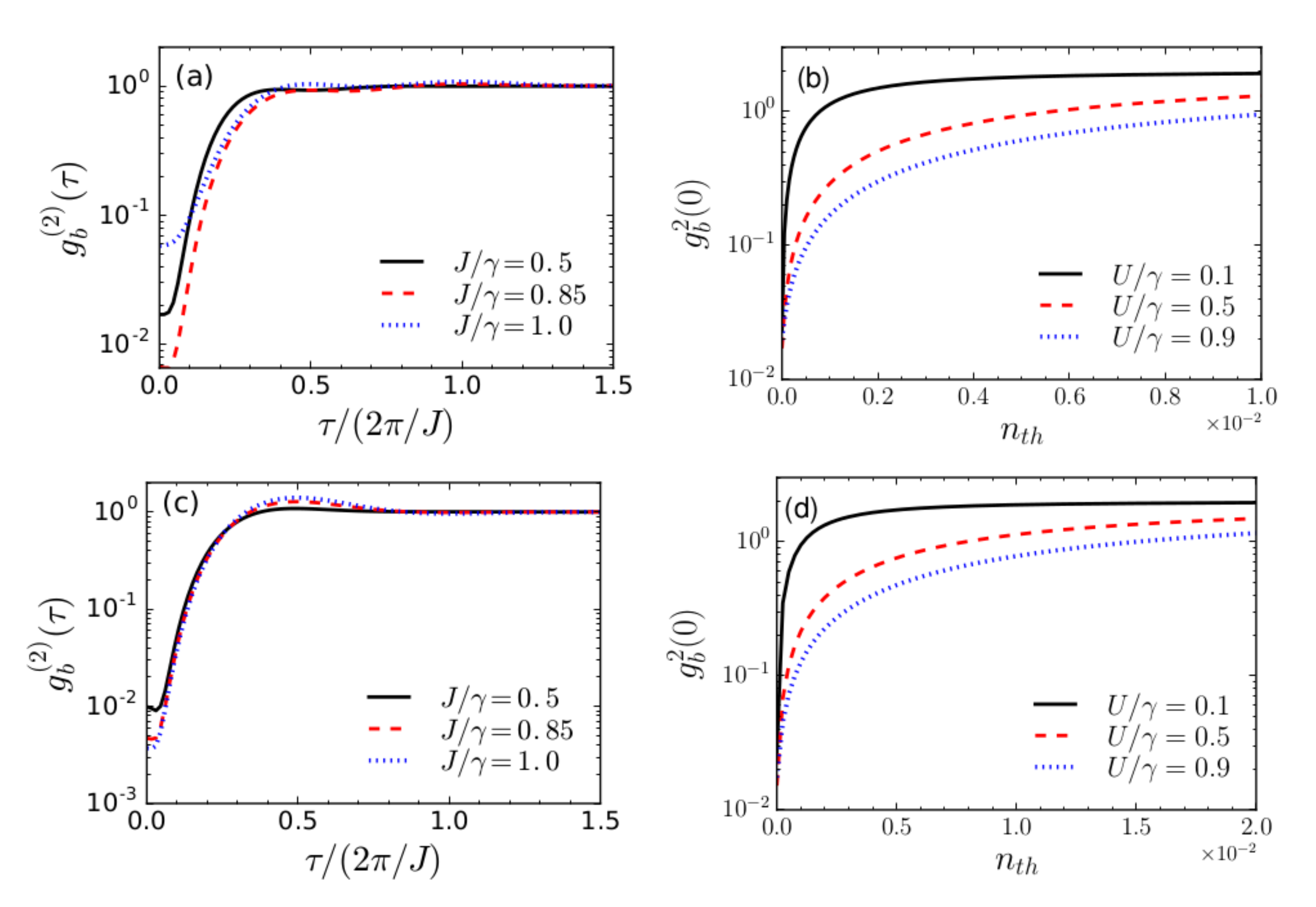}
	\caption {(a) Second-order correlation function with finite time-delay $g_b^{(2)}(\tau)$, and (b) effect of environmental temperature on $g_b^{(2)}(0)$ for $\zeta_+,\ \phi_+$. (c) Second-order correlation function with finite time-delay $g_b^{(2)}(\tau)$, and (d) effect of environmental temperature on $g_b^{(2)}(0)$ for $\zeta_-,\ \phi_-$. Other parameters are considered to be same as in Figs.~\ref{fig6}(c) and (d).}
	\label {fig7}
\end{figure}

Similarly, in Figs.~\ref{fig6}(a) and (b), we show $g_b^{(2)}(0)$ and the average phonon number in the primary resonator, as functions of the rescaled detuning $\Delta/\gamma$ and $U/\gamma$ corresponding to $\zeta_{-}, \phi_-$ for $J= 0.5\gamma$. Here also, $g_b^{(2)}(0)$ on the order of $0.01$ is obtained with average phonon number $\approx 0.01$. In Fig.~\ref{fig6}(c), we discuss the variation of $g_b^{(2)}(0)$ with respect to $\Delta/\gamma$ for $U_{\rm{opt}} = 0.5 \gamma$ and $\Delta_{\rm{opt}} = 0.15 \gamma$ and different values of $J/\gamma = 0.5$, $0.85$ and $1$. It is observed that phonon blockade can be obtained at $\Delta_{\rm{opt}} = 0.15 \gamma$. Fig.~\ref{fig6}(d) shows the variation of $g_b^{(2)}(0)$ for $U_{\rm{opt}} = 0.5 \gamma$ and $\Delta_{\rm{opt}} = -0.15 \gamma$. In this case, phonon blockade is obtained at $\Delta = -0.15 \gamma$.

Next, we discuss the variation of the second-order correlation function with finite time-delay, $g_b^{(2)}(\tau)$. In Figs.~\ref{fig7}(a) and (c), we show $g_b^{(2)}(\tau)$ as a function of the normalized time delay $\tau/(2\pi/J)$ with different values of $J$ for $\zeta_+, \phi_+$ and $\zeta_-, \phi_-$ respectively. It shows that the value of $g_b^{(2)}(0)$ is the lowest at $\tau = 0$ and for increasing delay times $g_b^{(2)}(\tau)>g_b^{(2)}(0)$, which demonstrates that the phonons are antibunched and sub-Poissonian in nature. In Figs.~\ref{fig7}(b) and (d), we discuss the effect of environmental phonon number on $g_b^{(2)}(0)$ for different values of $U/\gamma$, with $(\zeta_+, \phi_+$, $\Delta_{\textrm{opt}}/\gamma = 0.5$, $J/\gamma = 0.5)$ and $(\zeta_-, \phi_-$, $\Delta_{\textrm{opt}}/\gamma = 0.15$, $J/\gamma = 0.5)$ respectively. As observed in Fig.~\ref{fig7}(b), for optimum values of $\zeta_+, \phi_+$, the phonon blockade effect can be sustained upto $n_{\rm{th}} \approx 0.01$ for $U = 0.9 \gamma$ whereas for $U = 0.1 \gamma$ and $0.5 \gamma$, $g_b^{(2)}(0)\leq 1$ for values of $n_{\rm{th}}$ upto $\approx 0.001$ and $0.006$ respectively. On the other hand, for optimum values of $\zeta_-, \phi_-$, as shown in Fig.~\ref{fig7}(d), the phonon blockade effect can be sustained upto $n_{\rm{th}} \approx 0.02$ for $U = 0.9 \gamma$. For $U = 0.1 \gamma$ and $0.5 \gamma$, $g_b^{(2)}(0)\leq 1$ for $n_{\rm{th}}$ upto $\approx 0.001$ and $0.01$ respectively.

\subsection*{Measurement of phonon blockade via photon correlations}
\begin{figure}[!hbt]
	\centering
\includegraphics [width =0.8\linewidth]{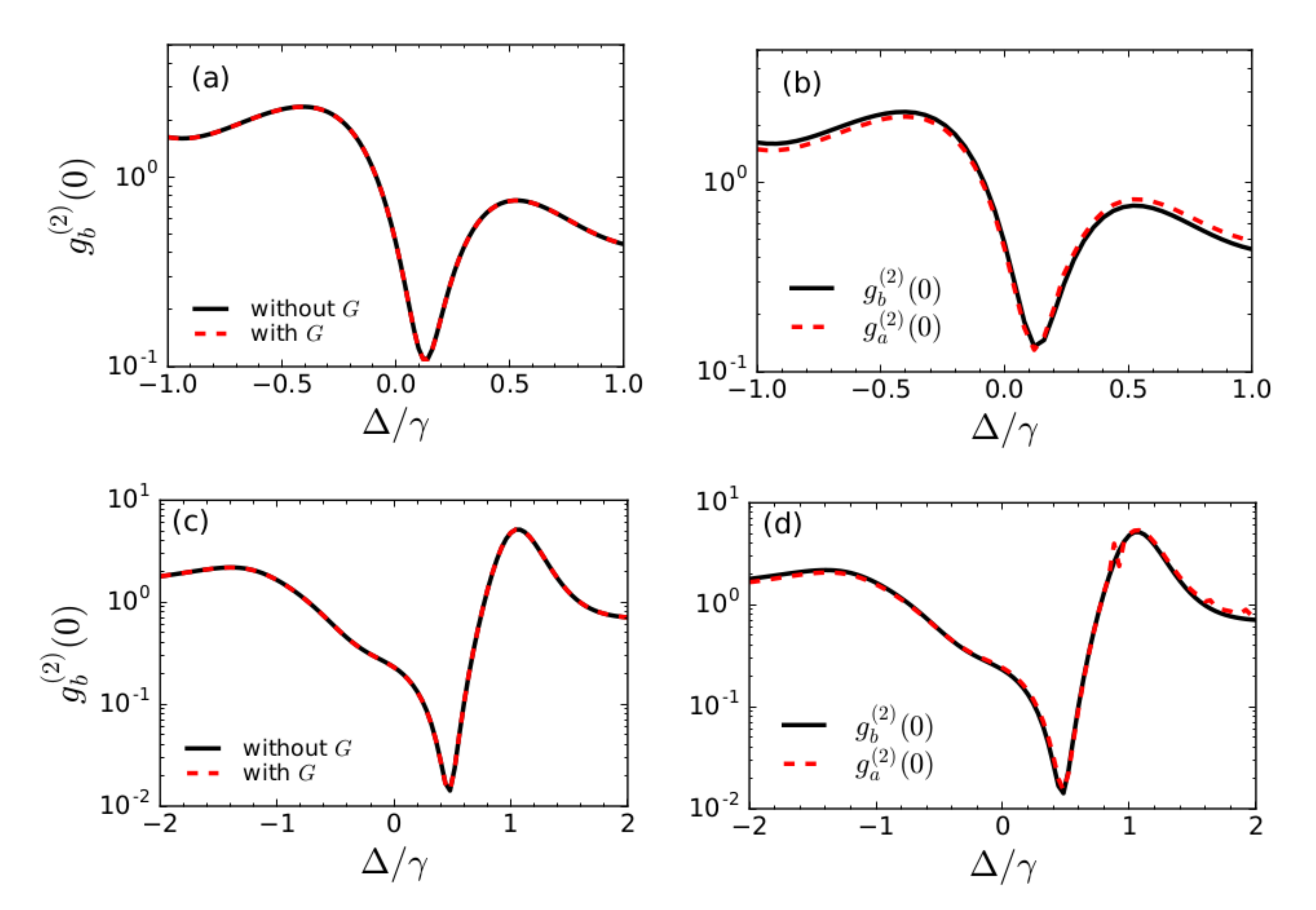}
	\caption {(a) The phonon correlations when the drive is applied only on the primary resonator i.e.~$\Omega_2 = 0$, showing $g_b^{(2)}(0)$ in absence of the optomechanical coupling (black solid line) and in presence of optomechanical coupling (red dashed line). (b) The phonon and photon correlations calculated for the total Hamiltonian in presence of the optomechanical coupling. (c) $g_b^{(2)}(0)$ in absence of the optomechanical coupling (black solid line) and in presence of optomechanical coupling (red dashed line) for additional driving of the secondary mode. (d) The phonon and photon correlations with two drives in presence of the optomechanical coupling. Other parameters are $\kappa=10 \gamma$, $G= 0.1\kappa$. }
	\label {fig8}
\end{figure}
Now, we study the phonon statistics in presence of the optomechanical interaction. Here, we will show that phonon blockade in the primary mechanical resonator can be detected by studying photon statistics of the optical mode in the cavity. Considering the cavity to be at the red sideband, the Langevin equation for the cavity mode fluctuation is given by
\begin{align} \label{eq11}
\dot{a}=-(i\omega_m+\frac{\kappa}{2}) -i G b_1 + \sqrt{\kappa}a_{\rm{in}}.
\end{align}
Here, $a_{in}$ is the input vacuum noise with the correlation function $\langle {a^\dagger}_{\rm{in}} (t) {a}_{\rm{in}} (t')\rangle\ =\ 0$. In the resolved sideband regime, i.e.~$\kappa\ll \omega_m$, and also for $\kappa \gg \{G, J, \gamma(n_{\rm{th}}+1)\}$, the cavity field follows the mechanical mode adiabatically \cite{ramos2013nonlinear, xu2015mechanical, xu2016phonon} 
\begin{align} \label{eqLangevin}
a = -i \frac{2G}{\kappa}b_1 + \rm{noise},
\end{align}
Therefore, $g_b^{(2)}(0) \approx g_a^{(2)}(0)=\langle a^\dagger a^\dagger a a\rangle/\langle a^\dagger a\rangle^2$ so that the phonon correlation can be studied by evaluating the second-order correlation function for photons. We calculate $g_b^{(2)}(0)$ and also the zero time-delay second-order correlation function for photon, $g_a^{(2)}(0)$, by solving the following master equation:
\begin{align} \label{eqmaster2}
{\dot{\rho}}_{tot}=& i[\rho_{tot}, H']+ \kappa L[a]\rho_{tot} +\gamma(n_{\textrm{th}}+1)L[b_1]\rho_{tot}+\gamma n_{\textrm{th}}L[b_1^\dagger]\rho_{tot} +\gamma (n_{\textrm{th}}+1)L[b_2]\rho_{tot}+\gamma n_{\textrm{th}}L[b_2^\dagger]\rho_{tot},
\end{align}
where the total Hamiltonian of the system, in a frame rotating at the mechanical pump frequency $\omega_p$ is given by
\begin{align} \label{eqH_prime_tot}
\nonumber
H'\ =\ & \Delta a^\dagger a + \Delta b_1^\dagger b_1+ \Delta b_2^\dagger b_2
+ U b_1^\dagger b_1^\dagger b_1 b_1 + U b_2^\dagger b_2^\dagger b_2 b_2+ J(b_1^\dagger b_2 + b_1 b_2^\dagger)+ G(a^\dagger b_1 + a b_1^\dagger)\\ & +\Omega_1(b_1^\dagger+b_1) + \Omega_2 (b_2^\dagger e^{-i\phi} +b_2 e^{i\phi}).
\end{align}
In Fig.~\ref{fig8}(a), we discuss the phonon correlations for only one drive applied on the primary resonator. We consider $G= 0.1\kappa$, that lies in the weak coupling regime and $\kappa=10 \gamma$ for typical optomechanical systems. The black solid line shows $g_b^{(2)}(0)$ in absence of the optomechanical coupling and the red dashed line shows the one in presence of the optomechanical coupling. It is observed that both the values agree well with each other in this parameter regime. Therefore, there is not any modification in the phonon blockade characteristics due to the additional coupling term induced by the optomechanical interaction in the adiabatic regime. In Fig.~\ref{fig8}(b), we compare the phonon and photon correlations calculated by solving the master equation for the total Hamiltonian in presence of the optomechanical coupling. We observe that both the correlation functions show evidence of blockade at the same detuning value. Therefore, the photon blockade characteristics for the cavity mode can serve as an evidence of phonon blockade in the primary mechanical resonator. Further, in Figs.~\ref{fig8}(c)-(d), we show $g_b^{(2)}(0)$ for additional driving of the secondary mode i.e~for $\Omega_2 \neq 0$, which also show similar features as the single-driving case.

\section*{Conclusion}
In conclusion, we have proposed schemes for the realization of phonon blockade in a weakly nonlinear mechanical end-mirror in an optomechanical cavity, coupled by Coulomb interaction to another weakly nonlinear mechanical resonator. Phonon correlations are characterised in terms of the second-order correlation function. Firstly, we studied the phonon blockade characteristics without considering the optomechanical interaction. By applying a single drive on the primary mechanical resonator, strong phonon blockade could be obtained with optimum values of the mechanical drive detuning and Kerr-nonlinearity. However, the phonon blockade effect is very fragile towards environmental thermal phonon number. Next, we discussed the scenario where both the mechanical resonators were driven simultaneously. In this case, the optimum values could be obtained in terms of the amplitude and the phase of the second mechanical drive, which allows more controllability of phonon blockade. Also, the phonon blockade effect could be sustained upto higher number of thermal phonons. Finally, we discussed the blockade characteristics to be observed when the optomechanical interaction was switched on. It was demonstrated that when the cavity optical field follows the resonator dynamics adiabatically, for both the single and the double mechanical drives, phonon blockade could be detected in terms of the photon correlations of the cavity mode.

\section*{Methods}
The optimal conditions for phonon blockade can be determined by solving the equations for the coefficients obtained from Schr\"{o}dinger equation: 
\begin{align} 
\label{eq19}
i\dot{C}_{10}=& \left(\Delta-i\frac{\gamma}{2}\right)C_{10}+JC_{01}+\Omega_1(C_{00}+\sqrt{2}C_{20}) +\Omega_2 e^{i\phi}C_{11},\\
\label{eq20}
i\dot{C}_{01} =& \left(\Delta-i\frac{\gamma}{2}\right)C_{01}+JC_{10}+\Omega_1C_{11} +\Omega_2 (e^{-i\phi}C_{00}+\sqrt{2}e^{i\phi}C_{02}),\\
\label{eq21}
i\dot{C}_{20} =& 2\left(\Delta + U-i\frac{\gamma}{2}\right) C_{20}+\sqrt{2}JC_{11}+\sqrt{2}\Omega_1 C_{10},\\
\label{eq22}
i\dot{C}_{11} =& 2\left(\Delta-i\frac{\gamma}{2}\right) C_{11}+\sqrt{2} J (C_{20}+C_{02}) +\Omega_1 C_{01}
+ \Omega_2 e^{-i\phi}C_{10},\\
\label{eq23}
i\dot{C}_{02} =& 2\left(\Delta + U-i\frac{\gamma}{2}\right) C_{02}+\sqrt{2}JC_{11}+\sqrt{2}\Omega_2 e^{-i\phi} C_{01}.
\end{align}
In the limit of weak $\Omega_1$ and $\Omega_2$, the probability of phonon excitation to higher levels becomes subsequently lower i.e.~$C_{00}\gg \{C_{10}, C_{01}\}\gg \{C_{20}, C_{11}, C_{02}\}$. The optimal condition for the complete phonon blockade in the primary resonator corresponds to the case when the probability of a phonon in state $|20\rangle$ equals zero. Under these assumptions, solving Eqs. \eqref{eq19} and \eqref{eq20}, the values of $C_{10}$ and $C_{01}$ at the steady-state are obtained as
\begin{align} 
\label{eq24}
C_{10}=& \frac{J\Omega_2 e^{-i \phi}-\Omega_1 (\Delta-i\frac{\gamma}{2})}{\left(\Delta-i\frac{\gamma}{2}\right)^2-J^2}C_{00},\\
\label{eq25}
C_{01}=& \frac{J\Omega_1 -\Omega_2 e^{-i \phi} (\Delta-i\frac{\gamma}{2})}{\left(\Delta-i\frac{\gamma}{2}\right)^2-J^2}C_{00}.
\end{align} 
Now, sustituting Eqs. \eqref{eq24} and \eqref{eq25} into Eqs. \eqref{eq21}-\eqref{eq23}, we obtain the following matrix equation
\begin{align}
\begin{pmatrix}
	x_{11}   & x_{12} & x_{13} \\
	x_{21}   & x_{22} & x_{23} \\
	x_{31}   & x_{32} & x_{33}
	\end{pmatrix}
	\begin{pmatrix}
	C_{11}\\
	C_{00}\\
	C_{02}
	 \end{pmatrix}
	 \ =\ 0,
\end{align} 
where, the matrix elements are given by
\begin{align}
\nonumber
x_{11}\ =&\ J,\quad 
x_{12}\ =\ \frac{J\Omega_1 \Omega_2 e^{-i \phi}-\Omega_1^2 (\Delta-i\frac{\gamma}{2})}{(\Delta-i\frac{\gamma}{2})^2-J^2},\quad
x_{13}\ =\ 0,\\
x_{21}\ =&\ 2(\Delta-i\frac{\gamma}{2}),\quad
x_{22}\ =\ \frac{J \left( \Omega_1^2 + \Omega_2^2 e^{-i \phi} \right)-2 \Omega_1 \Omega_2 e^{-i \phi} (\Delta-i\frac{\gamma}{2})}{(\Delta-i\frac{\gamma}{2})^2-J^2},\quad
x_{23}\ =\ \sqrt{2} J,\\
\nonumber
x_{31}\ =&\ J,\quad
x_{32}\ =\ \frac{J\Omega_1 \Omega_2 e^{-i \phi}-\Omega_2^2 e^{-2i \phi} (\Delta-i\frac{\gamma}{2})}{(\Delta-i\frac{\gamma}{2})^2-J^2},\quad
x_{33}\ =\ \sqrt{2} \left(\Delta + U -i\frac{\gamma}{2} \right).
\end{align}
To obtain nontrivial solutions for $C_{11}$, $C_{00}$ and $C_{02}$, the determinant of the coefficient matrix must be zero, from where we obtain the optimal parameters.



\begin{thebibliography}{10}
	\expandafter\ifx\csname url\endcsname\relax
	\def\url#1{\texttt{#1}}\fi
	\expandafter\ifx\csname urlprefix\endcsname\relax\def\urlprefix{URL }\fi
	\expandafter\ifx\csname doiprefix\endcsname\relax\def\doiprefix{DOI }\fi
	\providecommand{\bibinfo}[2]{#2}
	\providecommand{\eprint}[2][]{\url{#2}}
	
	\bibitem{caves1980measurement}
	\bibinfo{author}{Caves, C.~M.}, \bibinfo{author}{Thorne, K.~S.},
	\bibinfo{author}{Drever, R.~W.}, \bibinfo{author}{Sandberg, V.~D.} \&
	\bibinfo{author}{Zimmermann, M.}
	\newblock \bibinfo{journal}{\bibinfo{title}{On the measurement of a weak
			classical force coupled to a quantum-mechanical oscillator. {I}. issues of
			principle}}.
	\newblock {\emph{\JournalTitle{Rev. Mod. Phys.}}}
	\textbf{\bibinfo{volume}{52}}, \bibinfo{pages}{341} (\bibinfo{year}{1980}).
	
	\bibitem{bocko1996measurement}
	\bibinfo{author}{Bocko, M.~F.} \& \bibinfo{author}{Onofrio, R.}
	\newblock \bibinfo{journal}{\bibinfo{title}{On the measurement of a weak
			classical force coupled to a harmonic oscillator: experimental progress}}.
	\newblock {\emph{\JournalTitle{Rev. Mod. Phys.}}}
	\textbf{\bibinfo{volume}{68}}, \bibinfo{pages}{755} (\bibinfo{year}{1996}).
	
	\bibitem{nielsen2010quantum}
	\bibinfo{author}{Nielsen, M.~A.} \& \bibinfo{author}{Chuang, I.~L.}
	\newblock \emph{\bibinfo{title}{Quantum computation and quantum information}}
	(\bibinfo{publisher}{Cambridge university press}, \bibinfo{year}{2010}).
	
	\bibitem{stannigel2012optomechanical}
	\bibinfo{author}{Stannigel, K.} \emph{et~al.}
	\newblock \bibinfo{journal}{\bibinfo{title}{Optomechanical quantum information
			processing with photons and phonons}}.
	\newblock {\emph{\JournalTitle{Phys. Rev. Lett.}}}
	\textbf{\bibinfo{volume}{109}}, \bibinfo{pages}{013603}
	(\bibinfo{year}{2012}).
	
	\bibitem{rips2013quantum}
	\bibinfo{author}{Rips, S.} \& \bibinfo{author}{Hartmann, M.~J.}
	\newblock \bibinfo{journal}{\bibinfo{title}{Quantum information processing with
			nanomechanical qubits}}.
	\newblock {\emph{\JournalTitle{Phys. Rev. Lett.}}}
	\textbf{\bibinfo{volume}{110}}, \bibinfo{pages}{120503}
	(\bibinfo{year}{2013}).
	
	\bibitem{schwab2005putting}
	\bibinfo{author}{Schwab, K.~C.} \& \bibinfo{author}{Roukes, M.~L.}
	\newblock \bibinfo{journal}{\bibinfo{title}{Putting mechanics into quantum
			mechanics}}.
	\newblock {\emph{\JournalTitle{Phys. Today}}} \textbf{\bibinfo{volume}{58}},
	\bibinfo{pages}{36--42} (\bibinfo{year}{2005}).
	
	\bibitem{arcizet2006radiation}
	\bibinfo{author}{Arcizet, O.}, \bibinfo{author}{Cohadon, P.-F.},
	\bibinfo{author}{Briant, T.}, \bibinfo{author}{Pinard, M.} \&
	\bibinfo{author}{Heidmann, A.}
	\newblock \bibinfo{journal}{\bibinfo{title}{Radiation-pressure cooling and
			optomechanical instability of a micromirror}}.
	\newblock {\emph{\JournalTitle{Nature}}} \textbf{\bibinfo{volume}{444}},
	\bibinfo{pages}{71} (\bibinfo{year}{2006}).
	
	\bibitem{gigan2006self}
	\bibinfo{author}{Gigan, S.} \emph{et~al.}
	\newblock \bibinfo{journal}{\bibinfo{title}{Self-cooling of a micromirror by
			radiation pressure}}.
	\newblock {\emph{\JournalTitle{Nature}}} \textbf{\bibinfo{volume}{444}},
	\bibinfo{pages}{67} (\bibinfo{year}{2006}).
	
	\bibitem{kleckner2006sub}
	\bibinfo{author}{Kleckner, D.} \& \bibinfo{author}{Bouwmeester, D.}
	\newblock \bibinfo{journal}{\bibinfo{title}{Sub-kelvin optical cooling of a
			micromechanical resonator}}.
	\newblock {\emph{\JournalTitle{Nature}}} \textbf{\bibinfo{volume}{444}},
	\bibinfo{pages}{75} (\bibinfo{year}{2006}).
	
	\bibitem{teufel2009nanomechanical}
	\bibinfo{author}{Teufel, J.}, \bibinfo{author}{Donner, T.},
	\bibinfo{author}{Castellanos-Beltran, M.}, \bibinfo{author}{Harlow, J.} \&
	\bibinfo{author}{Lehnert, K.}
	\newblock \bibinfo{journal}{\bibinfo{title}{Nanomechanical motion measured with
			an imprecision below that at the standard quantum limit}}.
	\newblock {\emph{\JournalTitle{Nature Nanotech.}}}
	\textbf{\bibinfo{volume}{4}}, \bibinfo{pages}{820} (\bibinfo{year}{2009}).
	
	\bibitem{o2010quantum}
	\bibinfo{author}{O’Connell, A.~D.} \emph{et~al.}
	\newblock \bibinfo{journal}{\bibinfo{title}{Quantum ground state and
			single-phonon control of a mechanical resonator}}.
	\newblock {\emph{\JournalTitle{Nature}}} \textbf{\bibinfo{volume}{464}},
	\bibinfo{pages}{697} (\bibinfo{year}{2010}).
	
	\bibitem{habraken2012continuous}
	\bibinfo{author}{Habraken, S.}, \bibinfo{author}{Stannigel, K.},
	\bibinfo{author}{Lukin, M.~D.}, \bibinfo{author}{Zoller, P.} \&
	\bibinfo{author}{Rabl, P.}
	\newblock \bibinfo{journal}{\bibinfo{title}{Continuous mode cooling and phonon
			routers for phononic quantum networks}}.
	\newblock {\emph{\JournalTitle{New J. Phys.}}} \textbf{\bibinfo{volume}{14}},
	\bibinfo{pages}{115004} (\bibinfo{year}{2012}).
	
	\bibitem{gustafsson2014propagating}
	\bibinfo{author}{Gustafsson, M.~V.} \emph{et~al.}
	\newblock \bibinfo{journal}{\bibinfo{title}{Propagating phonons coupled to an
			artificial atom}}.
	\newblock {\emph{\it{Science}}} \textbf{\bibinfo{volume}{346}},
	\bibinfo{pages}{207--211} (\bibinfo{year}{2014}).
	
	\bibitem{kastner1993artificial}
	\bibinfo{author}{Kastner, M.~A.}
	\newblock \bibinfo{journal}{\bibinfo{title}{Artificial atoms}}.
	\newblock {\emph{\JournalTitle{Phys. Today}}} \textbf{\bibinfo{volume}{46}},
	\bibinfo{pages}{24--24} (\bibinfo{year}{1993}).
	
	\bibitem{imamoglu1997strongly}
	\bibinfo{author}{Imamoglu, A.}, \bibinfo{author}{Schmidt, H.},
	\bibinfo{author}{Woods, G.} \& \bibinfo{author}{Deutsch, M.}
	\newblock \bibinfo{journal}{\bibinfo{title}{Strongly interacting photons in a
			nonlinear cavity}}.
	\newblock {\emph{\JournalTitle{Phys. Rev. Lett.}}}
	\textbf{\bibinfo{volume}{79}}, \bibinfo{pages}{1467} (\bibinfo{year}{1997}).
	
	\bibitem{liu2010qubit}
	\bibinfo{author}{Liu, Y.-X.} \emph{et~al.}
	\newblock \bibinfo{journal}{\bibinfo{title}{Qubit-induced phonon blockade as a
			signature of quantum behavior in nanomechanical resonators}}.
	\newblock {\emph{\JournalTitle{Phys. Rev. A}}} \textbf{\bibinfo{volume}{82}},
	\bibinfo{pages}{032101} (\bibinfo{year}{2010}).
	
	\bibitem{miranowicz2016tunable}
	\bibinfo{author}{Miranowicz, A.}, \bibinfo{author}{Bajer, J.},
	\bibinfo{author}{Lambert, N.}, \bibinfo{author}{Liu, Y.-X.} \&
	\bibinfo{author}{Nori, F.}
	\newblock \bibinfo{journal}{\bibinfo{title}{Tunable multiphonon blockade in
			coupled nanomechanical resonators}}.
	\newblock {\emph{\JournalTitle{Phys. Rev. A}}} \textbf{\bibinfo{volume}{93}},
	\bibinfo{pages}{013808} (\bibinfo{year}{2016}).
	
	\bibitem{wang2016method}
	\bibinfo{author}{Wang, X.}, \bibinfo{author}{Miranowicz, A.},
	\bibinfo{author}{Li, H.-R.} \& \bibinfo{author}{Nori, F.}
	\newblock \bibinfo{journal}{\bibinfo{title}{Method for observing robust and
			tunable phonon blockade in a nanomechanical resonator coupled to a charge
			qubit}}.
	\newblock {\emph{\JournalTitle{Phys. Rev. A}}} \textbf{\bibinfo{volume}{93}},
	\bibinfo{pages}{063861} (\bibinfo{year}{2016}).
	
	\bibitem{ramos2013nonlinear}
	\bibinfo{author}{Ramos, T.}, \bibinfo{author}{Sudhir, V.},
	\bibinfo{author}{Stannigel, K.}, \bibinfo{author}{Zoller, P.} \&
	\bibinfo{author}{Kippenberg, T.~J.}
	\newblock \bibinfo{journal}{\bibinfo{title}{Nonlinear quantum optomechanics via
			individual intrinsic two-level defects}}.
	\newblock {\emph{\JournalTitle{Phys. Rev. Lett.}}}
	\textbf{\bibinfo{volume}{110}}, \bibinfo{pages}{193602}
	(\bibinfo{year}{2013}).
	
	\bibitem{xie2017phonon}
	\bibinfo{author}{Xie, H.}, \bibinfo{author}{Liao, C.-G.},
	\bibinfo{author}{Shang, X.}, \bibinfo{author}{Ye, M.-Y.} \&
	\bibinfo{author}{Lin, X.-M.}
	\newblock \bibinfo{journal}{\bibinfo{title}{Phonon blockade in a quadratically
			coupled optomechanical system}}.
	\newblock {\emph{\JournalTitle{Phys. Rev. A}}} \textbf{\bibinfo{volume}{96}},
	\bibinfo{pages}{013861} (\bibinfo{year}{2017}).
	
	\bibitem{seok2017antibunching}
	\bibinfo{author}{Seok, H.} \& \bibinfo{author}{Wright, E.}
	\newblock \bibinfo{journal}{\bibinfo{title}{Antibunching in an optomechanical
			oscillator}}.
	\newblock {\emph{\JournalTitle{Phys. Rev. A}}} \textbf{\bibinfo{volume}{95}},
	\bibinfo{pages}{053844} (\bibinfo{year}{2017}).
	
	\bibitem{xu2016phonon}
	\bibinfo{author}{Xu, X.-W.}, \bibinfo{author}{Chen, A.-X.} \&
	\bibinfo{author}{Liu, Y.-X.}
	\newblock \bibinfo{journal}{\bibinfo{title}{Phonon blockade in a nanomechanical
			resonator resonantly coupled to a qubit}}.
	\newblock {\emph{\JournalTitle{Phys. Rev. A}}} \textbf{\bibinfo{volume}{94}},
	\bibinfo{pages}{063853} (\bibinfo{year}{2016}).
	
	\bibitem{guan2017phonon}
	\bibinfo{author}{Guan, S.}, \bibinfo{author}{Bowen, W.~P.},
	\bibinfo{author}{Liu, C.} \& \bibinfo{author}{Duan, Z.}
	\newblock \bibinfo{journal}{\bibinfo{title}{Phonon antibunching effect in
			coupled nonlinear micro/nanomechanical resonator at finite temperature}}.
	\newblock {\emph{\JournalTitle{EPL}}} \textbf{\bibinfo{volume}{119}},
	\bibinfo{pages}{58001} (\bibinfo{year}{2017}).
	
	\bibitem{shi2018tunable}
	\bibinfo{author}{Shi, H.-Q.}, \bibinfo{author}{Zhou, X.-T.},
	\bibinfo{author}{Xu, X.-W.} \& \bibinfo{author}{Liu, N.-H.}
	\newblock \bibinfo{journal}{\bibinfo{title}{Tunable phonon blockade in
			quadratically coupled optomechanical systems}}.
	\newblock {\emph{\JournalTitle{Sci. Rep.}}} \textbf{\bibinfo{volume}{8}},
	\bibinfo{pages}{2212} (\bibinfo{year}{2018}).
	
	\bibitem{liew2010single}
	\bibinfo{author}{Liew, T. C.~H.} \& \bibinfo{author}{Savona, V.}
	\newblock \bibinfo{journal}{\bibinfo{title}{Single photons from coupled quantum
			modes}}.
	\newblock {\emph{\JournalTitle{Phys. Rev. Lett.}}}
	\textbf{\bibinfo{volume}{104}}, \bibinfo{pages}{183601}
	(\bibinfo{year}{2010}).
	
	\bibitem{bamba2011origin}
	\bibinfo{author}{Bamba, M.}, \bibinfo{author}{Imamoglu, A.},
	\bibinfo{author}{Carusotto, I.} \& \bibinfo{author}{Ciuti, C.}
	\newblock \bibinfo{journal}{\bibinfo{title}{Origin of strong photon
			antibunching in weakly nonlinear photonic molecules}}.
	\newblock {\emph{\JournalTitle{Phys. Rev. A}}} \textbf{\bibinfo{volume}{83}},
	\bibinfo{pages}{021802} (\bibinfo{year}{2011}).
	
	\bibitem{tang2015quantum}
	\bibinfo{author}{Tang, J.}, \bibinfo{author}{Geng, W.} \& \bibinfo{author}{Xu,
		X.}
	\newblock \bibinfo{journal}{\bibinfo{title}{Quantum interference induced photon
			blockade in a coupled single quantum dot-cavity system}}.
	\newblock {\emph{\JournalTitle{Sci. Rep.}}} \textbf{\bibinfo{volume}{5}}
	(\bibinfo{year}{2015}).
	
	\bibitem{sarma2017quantum}
	\bibinfo{author}{Sarma, B.} \& \bibinfo{author}{Sarma, A.~K.}
	\newblock \bibinfo{journal}{\bibinfo{title}{Quantum-interference-assisted
			photon blockade in a cavity via parametric interactions}}.
	\newblock {\emph{\JournalTitle{Phys. Rev. A}}} \textbf{\bibinfo{volume}{96}},
	\bibinfo{pages}{053827} (\bibinfo{year}{2017}).
	
	\bibitem{flayac2017unconventional}
	\bibinfo{author}{Flayac, H.} \& \bibinfo{author}{Savona, V.}
	\newblock \bibinfo{journal}{\bibinfo{title}{Unconventional photon blockade}}.
	\newblock {\emph{\JournalTitle{Phys. Rev. A}}} \textbf{\bibinfo{volume}{96}},
	\bibinfo{pages}{053810} (\bibinfo{year}{2017}).
	
	\bibitem{didier2011detecting}
	\bibinfo{author}{Didier, N.}, \bibinfo{author}{Pugnetti, S.},
	\bibinfo{author}{Blanter, Y.~M.} \& \bibinfo{author}{Fazio, R.}
	\newblock \bibinfo{journal}{\bibinfo{title}{Detecting phonon blockade with
			photons}}.
	\newblock {\emph{\JournalTitle{Phys. Rev. B}}} \textbf{\bibinfo{volume}{84}},
	\bibinfo{pages}{054503} (\bibinfo{year}{2011}).
	
	\bibitem{hensinger2005ion}
	\bibinfo{author}{Hensinger, W.} \emph{et~al.}
	\newblock \bibinfo{journal}{\bibinfo{title}{Ion trap transducers for quantum
			electromechanical oscillators}}.
	\newblock {\emph{\JournalTitle{Phys. Rev. A}}} \textbf{\bibinfo{volume}{72}},
	\bibinfo{pages}{041405} (\bibinfo{year}{2005}).
	
	\bibitem{ma2014tunable}
	\bibinfo{author}{Ma, P.-C.}, \bibinfo{author}{Zhang, J.-Q.},
	\bibinfo{author}{Xiao, Y.}, \bibinfo{author}{Feng, M.} \&
	\bibinfo{author}{Zhang, Z.-M.}
	\newblock \bibinfo{journal}{\bibinfo{title}{Tunable double optomechanically
			induced transparency in an optomechanical system}}.
	\newblock {\emph{\JournalTitle{Phys. Rev. A}}} \textbf{\bibinfo{volume}{90}},
	\bibinfo{pages}{043825} (\bibinfo{year}{2014}).
	
	\bibitem{chen2015dissipation}
	\bibinfo{author}{Chen, R.-X.}, \bibinfo{author}{Shen, L.-T.} \&
	\bibinfo{author}{Zheng, S.-B.}
	\newblock \bibinfo{journal}{\bibinfo{title}{Dissipation-induced optomechanical
			entanglement with the assistance of coulomb interaction}}.
	\newblock {\emph{\JournalTitle{Phys. Rev. A}}} \textbf{\bibinfo{volume}{91}},
	\bibinfo{pages}{022326} (\bibinfo{year}{2015}).
	
	\bibitem{xu2015mechanical}
	\bibinfo{author}{Xu, X.-W.}, \bibinfo{author}{Liu, Y.-X.},
	\bibinfo{author}{Sun, C.-P.} \& \bibinfo{author}{Li, Y.}
	\newblock \bibinfo{journal}{\bibinfo{title}{Mechanical {PT} symmetry in coupled
			optomechanical systems}}.
	\newblock {\emph{\JournalTitle{Phys. Rev. A}}} \textbf{\bibinfo{volume}{92}},
	\bibinfo{pages}{013852} (\bibinfo{year}{2015}).
	
\end{thebibliography}


\end{document}